\documentclass{article}
\usepackage[utf8]{inputenc}
\usepackage{authblk}
\usepackage[dvips]{graphicx}
\graphicspath{{noiseimages/}}
\usepackage{xcolor}
\usepackage{float}
\usepackage[charter]{mathdesign}
\let\mathcal\undefined
\DeclareMathAlphabet{\mathcal}{OMS}{cmsy}{m}{n}
\usepackage[T1]{fontenc}
\usepackage{tikz}
\usepackage{empheq}

\usepackage{MnSymbol}
\usepackage{pgfplots}
\pgfplotsset{compat=1.7}
\usetikzlibrary{intersections, pgfplots.fillbetween}
\usetikzlibrary{decorations.pathmorphing}
\usetikzlibrary{arrows.meta}
\usepackage[mode=buildnew]{standalone}
\usepackage[toc,page]{appendix}
\usepackage{amsmath}
\usepackage{float}
\usepackage{comment}
\usepackage{bm}
\usepackage{a4wide}
\usepackage{bm}

\usepackage{cite}

\usepackage[english]{babel}
\usepackage{hyperref}
\usepackage{caption}
\AtBeginDocument{}
\definecolor{refcolor}{HTML}{CD2600}
\definecolor{tablecolor}{HTML}{373641} 
\definecolor{urlcolor}{HTML}{E12900} 
\hypersetup{pdfstartview=FitH,  linkcolor=refcolor,urlcolor=urlcolor, colorlinks=true,citecolor=refcolor}
\usepackage[page]{appendix}

\author[1,2]{E. T. Akhmedov}
\author[1,2]{D. V. Diakonov}
\author[3]{C. Schubert}
\affil[1]{Moscow Institute of Physics and Technology, Institutskii per. 9, 141700, Dolgoprudny, Russia}
\affil[2]{NRC ``Kurchatov Institute'', 123182, Moscow, Russia}
\affil[3]{Facultad de Ciancias Físico-Matemáticas, Universidad Michoacan de San Nicolás de Hidalgo, Morelia, Mexico}
\title{\textcolor{black}{Complex effective actions and gravitational pair creation}}
\begin{document}

\numberwithin{equation}{section}

\maketitle

\begin{abstract}
    We use different methods to calculate the imaginary part of the gravitational effective action due to a massless scalar field, with a view on perturbative vs. non-pertubative and the results of ArXiv:2305.18521.
\end{abstract}

\newpage 

\section{Introduction}

We consider the free (gaussian) real scalar field theory with an arbitrary nonminimal coupling, with the parameter $\xi$, to the curvature, $R$:
\begin{align}
    S[g_{\mu \nu},\phi]=-\frac{1}{2}\int d^{4}x \, \sqrt{g} \, \Big[g^{\mu \nu} \partial_\mu \phi(x)\partial_\nu \phi(x)+m^2\phi^2(x)+\xi R \phi^2(x)\Big],
\end{align}
where $g = |\det g_{\mu\nu}|$.
In the gaussian approximation the one loop effective action is defined as follows:
\begin{align}
     \langle out |in\rangle=e^{i W}=\int [d \phi]e^{i S}.
\end{align}
According to Schwinger's general definition the probability that at least one particle-antiparticle pair is created by the gravitational field out of the initial Fock space ground state is given by the imaginary part of the effective action:
\begin{align}
    P=1-|\langle out |in\rangle|^2=1-e^{-2 Im\, W}.
\end{align}
For small value of the particle creation rate we have: 
\begin{align}
    P\approx 2 Im\, W .
\end{align} 
Let us summarize several different methods to find the imaginary part of the one loop effective action that we describe below in grater details:
\begin{itemize}
    \item For the massless conformaly coupled, 
    $\left(\xi=\frac{1}{6}\right)$, scalar field one can use the expansion up to the second order in the pertubation of the metric, $h$. Then using the general covariance one can restore the whole effective action, including its imaginary part \cite{Frieman:1985fr,Campos:1993ug,Elias:2017wkr}:
\begin{gather}
    Im\, W = \raisebox{-.4\height}{\includegraphics[scale = 0.9]{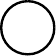}} \quad+ \quad \raisebox{-.4\height}{\includegraphics[scale = 0.9]{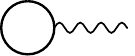}}+\raisebox{-.4\height}{\includegraphics[scale = 0.9]{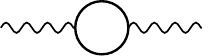}} \quad+ \quad \raisebox{-.4\height}{\includegraphics[scale = 0.9]{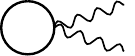}}
\nonumber
= \\=
\frac{1}{2}\frac{1}{2880
    \pi}\int \frac{d^4p}{(2\pi)^4} \theta(-p^2) \left( R^{(2)}_{\mu\nu\alpha\beta}(p)R^{(2) \mu\nu\alpha\beta}(-p)-R_{\mu\nu}^{(2)}(p)R^{(2)\mu\nu}(-p)\right)+O(h^3).
\end{gather}
After that, using the vanishing of the Gauss-Bonnet topological invariant the above expression can be rewritten in terms of the Weyl tensor:
\begin{align}
  Im\, W = \frac{1}{2}\frac{1}{1920\pi} \int \frac{d^4 p}{(2\pi)^4} \theta(-p^2) C^{(2)}_{\mu\nu\alpha\beta}(p) C^{(2)\mu\nu\alpha\beta}(-p) +O(h^3).
\end{align}
The whole calculation is performed in the Lorentzian signature. 

\item Using the heat kernel expansion of the effective action up to higher order terms in the curvature tensor the imaginary part of the effective action can be shown to be equal to \cite{Dobado:1998mr}: 
\begin{gather}
  Im\, W = \\=\nonumber \frac{1}{2}\frac{1}{1920\pi} \int \frac{d^4 p}{(2\pi)^4} \theta(-p^2) \left[C_{\mu\nu\alpha\beta}(p) C^{\mu\nu\alpha\beta}(-p) + 60 \left(\xi-\frac{1}{6}\right)^2 R(p)R(-p)\right]+O(R^3).
\end{gather}
It reduces to the previous answer for the case of the conformal coupling, $\left(\xi=\frac{1}{6}\right)$.

The FLRW metric is locally conformal to the Minkowski metric and, hence, its Weyl tensor vanishes. Hence the imaginary part of effective action is given by:
\begin{gather}
  Im\, W =  \frac{1}{64\pi} \int d^4 x \sqrt{g}    \left(\xi-\frac{1}{6}\right)^2 R(x)^2+O(R^3),
\end{gather}
after the inverse Fourier transformation. This result is only valid for homogeneous and isotropic metrics.

\item In \cite{Wondrak:2023zdi} the effective action in the Euclidean signature is found using the heat kernel expansion. They obtain the following ``exact'' expression for the imaginary part of the effective action:  
\begin{gather}
  Im\, W = \\=\nonumber \frac{1}{2}\Bigg[\frac{1}{2}\frac{1}{1920\pi} \int d^4 x \sqrt{g} \left[C_{\mu\nu\alpha\beta}(x) C^{\mu\nu\alpha\beta}(x) + 60 \left(\xi-\frac{1}{6}\right)^2 R^2(x)\right]\Bigg].
\end{gather}
Which is two times smaller than the results found above. The origin of this discrepancy is not yet clear for us, but we think that the two results are ought to match.
\end{itemize}

There are also non-perturbative methods of computation of the imaginary part of the one loop effective action, which are generally based either on the formalism of Bogoliubov transformations \cite{ Parker:1968mv, Parker:1969au, Birrell:1982ix, Mottola:1984ar,Anderson:2013ila, Ford:2021syk} or the relation of the effective action to the coincident limit of the Feynman In-Out propagator
\cite{Candelas:1975du,Das:2006wg,Polyakov:2007mm,Akhmedov:2009ta,Akhmedov:2019esv,Akhmedov:2021agm}.The latter methods are related to the ones listed above, but will not be discussed in the present paper. However, it is important to mention that, for de Sitter space and in the large mass/weak-field approximation, they yield the non-perturbative result
\cite{Akhmedov:2019esv}

\begin{eqnarray}
{\rm Im} L \approx \frac{1}{8\pi^2} m^{\frac{3}{2}}  \,{\rm e}^{-2\pi m}
\end{eqnarray}
which is presumably the closest possible analogue to 
Schwinger's QED formula for a constant electric field,

\begin{eqnarray}
Im \, L_{EH}(E) = \frac{m^4}{8\pi^3}
\beta^2\, \sum_{k=1}^\infty \frac{1}{k^2}
\,\exp\left[-\frac{\pi k}{\beta}\right]
\label{schwinger}
\end{eqnarray}
($\beta = eE/m^2$).

The present work was triggered by the claims of \cite{Wondrak:2023zdi} (see also the the comment by \cite{Ferreiro:2023jfs}).  
Its main purpose is to discuss the proper distinction betweem
perturbative vs. non-perturbative pair creation, and to point out some peculiarities of the massless case.

\section{The expansion over the metric perturbations}

In this section we will present the method based on the expansion over the metric perturbations to find the imaginary part of the one loop effective action.  The action of the free (gaussian) massless conformally coupled scalar field $\phi$ in a curved space-time is given by:

\begin{align}
    S[g_{\mu \nu},\phi]=-\frac{1}{2}\int d^{d}x \, \sqrt{g} \,  \Big[g^{\mu \nu} \partial_\mu \phi(x)\partial_\nu \phi(x)+\xi R \phi^2(x)\Big],
\end{align}
where $d$ is the number of space-time dimensions and $\xi=\frac{d-2}{4(d-1)}$ for the case of the conformal coupling.

Using the perturbative expansion of the metric around the Minkowskian background, $\eta_{\mu\nu}$:
\begin{align}
    g_{\mu \nu}(x)=\eta_{\mu \nu}+h_{\mu \nu}(x),
\end{align}
where $h_{\mu\nu}$ is a perturbation, the above action can be expanded as follows: 
\begin{align}
    S[g_{\mu \nu},\phi]=S^0[ \eta_{\mu \nu},\phi]+\sum_{n=1}^\infty S^{n}_I[h_{\mu \nu},\phi].
\end{align}
Here $S^0$ is the action for the field in flat spacetime, and $S^{n}_I$ is the n-th order expansion in $h$.  

The zero and the first order terms are defined as follows:
\begin{align}
    S^0[ \eta_{\mu \nu},\phi]=-\frac{1}{2}\int d^d x \Big[\eta^{\mu\nu} \partial_\mu \phi(x)\partial_\nu \phi(x)\Big]
\end{align}
and
\begin{align}
   S^{1}_I[h_{\mu \nu},\phi] =\frac{1}{2}\int d^d x \Bigg[\left(h^{\mu\nu}-\frac{1}{2}h \eta^{\mu\nu}\right) \partial_\mu \phi(x)\partial_\nu \phi(x)-\xi \left(\partial_\mu \partial_\nu h^{\mu\nu}-\partial_\mu \partial^\mu h\right) \phi^2(x)\Bigg]. 
\end{align}
Here we use the series expansion: 
\begin{align}
    g^{\mu \nu}(x)=\eta^{\mu \nu}-h^{\mu \nu}+O(h^2),
\end{align}
\begin{align}
    \sqrt{g}=1+\frac{1}{2}h+O(h^2)
\end{align}
and 
\begin{align}
    R=\partial_\mu \partial_\nu h^{\mu\nu}-\partial_\mu \partial^\mu h+O(h^2).
\end{align}
The effective gravitational action, after the integration over the matter field, is defined as follows:
\begin{align}
    e^{i W}=\int d[ \phi] e^{i S}.
\end{align}
It can be expanded as the sum of all connected Feynman diagrams: 
\begin{align}
    W = \sum_{n=0}^\infty W_{n},
\end{align}
where index $n$ defines n-th order expansion in $h$. 

To second order the graphical representation of the effective action is as follows:
\begin{align}
    i W_0=\log\int d [\phi] e^{i S_0}=\quad \raisebox{-.4\height}{\includegraphics[scale = 0.9]{ef_circle.pdf}},
\end{align}
\begin{align}
    i W_1=\frac{\int d [\phi] e^{i S_0} i S_I^{1}}{\int d[\phi] e^{i S_0}}= \quad \raisebox{-.4\height}{\includegraphics[scale = 0.9]{ef_circle1.pdf}},
\end{align}
and 
\begin{gather}
     i W_2=-\frac{1}{2}\frac{\int d[\phi] e^{i S_0}  \left(S_I^{1}\right)^2}{\int d[\phi] e^{i S_0}}+\frac{\int d[\phi] e^{i S_0} i S_I^{2}}{\int d [\phi] e^{i S_0}} 
     =\\= \quad
     \nonumber
     \raisebox{-.4\height}{\includegraphics[scale = 0.9]{ef_circle2.pdf}} \quad+ \quad \raisebox{-.4\height}{\includegraphics[scale = 0.9]{ef_circle3.pdf}}.
\end{gather}
The terms $W_0$, $W_1$ and the second term in $W_2$ vanish in the dimensional regularization, since the integrals of the following form are equal to zero:
\begin{align}
    \int d^d k \frac{1}{k^2}=0, \quad \int d^d k \frac{k_\mu}{k^2}=0,\quad \text{and} \quad  \int d^d k \frac{k_\mu k_\nu}{k^2}=0.
\end{align}
Therefore the first non vanishing contribution to the effective action appears at the second order in $h$ and is equal to: 
\begin{gather}
\nonumber
     W = \frac{i}{2}\frac{\int d[\phi] e^{i S_0}  \left(S_I^{1}\right)^2}{\int d[\phi] e^{i S_0}}+O(h^3)
     = \quad
     \raisebox{-.4\height}{\includegraphics[scale = 0.9]{ef_circle2.pdf}} \quad = \\= 
     \label{W}
     -\frac{1}{4} \frac{1}{2880 \pi^2}\Bigg[ \frac{1}{d-4}\int d^4 x\left[3 \, R^{(2)}_{\mu\nu\alpha\beta}(x)R^{(2) \mu\nu\alpha\beta}(x)-(R^{(2)})^2\right] +\frac{1}{3} \int d^4 x \, R^{(2)}(x)R^{(2)}(x) 
     -\\-
     \nonumber
     \int d^4x d^4 y \left[3 R^{(2)}_{\mu\nu\alpha\beta}(x)R^{(2) \mu\nu\alpha\beta}(y)-R^{(2)}(x)R^{(2)}(y)\right] \, K(x-y)+(d-4)\Bigg],
     \end{gather}
where $R^{(2)}_{\mu\nu\alpha\beta}(x)$ is the Riemann tensor and $R^{(2)}(x)$ is the Ricci scalar expanded up to the second order in $h$, and $K(x-y)$ is defined as follows:
\begin{align}
    K(x-y)=-\frac{1}{2}\int \frac{d^4 p}{(2\pi)^4}e^{i p \cdot(x-y)} \log\left[\frac{p^2-i \epsilon}{\mu^2}\right].
\end{align}
The imaginary part comes only from the term on the second line in \eqref{W}.  In fact, the imaginary part of $K(x-y)$ is equal to:
\begin{align}
   Im \, K(x-y)=\frac{\pi}{2}\int \frac{d^4 p}{(2\pi)^4}e^{i p \cdot(x-y)} \theta(-p^2).
\end{align}
Therefore the imaginary part of the effective action up to the second order in $h$ is given by:
\begin{align}
    P=2 Im\, W= \frac{1}{4}\frac{1}{2880
    \pi}\int \frac{d^4p}{(2\pi)^4} \theta(-p^2) \left[3 \,  R^{(2)}_{\mu\nu\alpha\beta}(p)R^{(2) \mu\nu\alpha\beta}(-p)-R^{(2)}(p)R^{(2)}(-p)\right].
\end{align}
The curvature-squared terms are not independent since one can add the Gauss-Bonnet term. Namely, in the second order in $h$ \cite{Frieman:1985fr}:
\begin{align}
\label{curvature-squared terms are not independent}
    R^{(2)}_{\mu\nu\alpha\beta}(p) R^{(2)\mu\nu\alpha\beta}(-p) -4   R^{(2)}_{\mu\nu}(p) R^{(2)\mu\nu}(-p) +R^{(2)}(p)R^{(2)}(-p)=0.
\end{align}
Then, one can rewrite the imaginary part of the effective ection in the following form:
\begin{align}
\label{p}
   Im\, W =\frac{1}{2} \frac{1}{2880
    \pi}\int \frac{d^4p}{(2\pi)^4} \theta(-p^2) \left( R^{(2)}_{\mu\nu\alpha\beta}(p)R^{(2) \mu\nu\alpha\beta}(-p)-R_{\mu\nu}^{(2)}(p)R^{(2)\mu\nu}(-p)\right)
\end{align}
or in terms of the Weyl tensor:
\begin{align}
\label{P1}
   Im\, W =\frac{1}{2} \frac{1}{1920\pi} \int \frac{d^4 p}{(2\pi)^4} \theta(-p^2) C^{(2)}_{\mu\nu\alpha\beta}(p) C^{(2)\mu\nu\alpha\beta}(-p) ,
\end{align}
where: 
\begin{align}
    C^{(2)}_{\mu\nu\alpha\beta}(p) C^{(2)\mu\nu\alpha\beta}(-p)=  R^{(2)}_{\mu\nu\alpha\beta}(p) R^{(2)\mu\nu\alpha\beta}(-p) -2   R^{(2)}_{\mu\nu}(p) R^{(2)\mu\nu}(-p) +\frac{1}{3} R^{(2)}(p)R^{(2)}(-p).
\end{align}
If one will analytically continue this expression to the Euclidean signature one will find that $\theta(-p^2) = 1$ under the integrals in \eqref{p} and \eqref{P1}. After the analytical continuation one will get the following expression after the inverse Fourier transformation of the fields:
\begin{align}
  Im\, W =\frac{1}{2} \frac{1}{1920\pi} \int d^4 x  C^{(2)}_{\mu\nu\alpha\beta}(x) C^{(2)\mu\nu\alpha\beta}(x),
\end{align}
 in agreement with the expressions computed by other means \cite{Frieman:1985fr,Campos:1993ug}. 
 
From the obtained expression we can restore the answer to all orders in $h$ using just the general covariance:
 \begin{align}
 \label{WperC}
   Im\, W =\frac{1}{2} \frac{1}{1920\pi} \int d^4x \sqrt{g}  C_{\mu\nu\alpha\beta}(x) C^{\mu\nu\alpha\beta}(x).
\end{align}
From the derivation that has been used above one can see that this is the imaginary part to the lowest order in the curvature tensor. In fact, there also will be higher terms in the curvature.

\section{The heat kernel expansion method}

In this section we will use the nonlocal Barvinsky-Vilkovisky form-factor representation of the effective action to express its imaginary part up to higher order terms in curvature. 

The one-loop effective action can be expressed as:
 \begin{align}
     W=-\frac{1}{2}\int_0^\infty \frac{ds}{s} e^{-i s m^2 } Tr \  K(s),
 \end{align}
where the heat kernel is defined as follows \cite{Barvinsky:1990up}: 
     \begin{gather}
         Tr \, K(s)=Tr e^{-i s\left(-\Box+\xi R-i\epsilon\right)}=\frac{1}{(4\pi s)^{d/2}} \int d^d x \sqrt{g}\Bigg[1+i s  \left(\frac{1}{6}-\xi\right)R
+\\+ 
\nonumber
(is)^2\left[R f_R(s \Box) R+ R_{\mu\nu\alpha\beta}(x)f_4(s \Box) R^{\mu\nu\alpha\beta}(x) -  R_{\mu\nu}(x)f_2(s\Box) R^{\mu\nu}(x)\right]\Bigg]+...
     \end{gather}
where the exact definition of the functions $f_i(s \Box)$ can be found in \cite{Barvinsky:1990up}.
    
More specifically, the result for $W$ in the massless scalar field theory to the second order in the curvature can be written as: 
\begin{gather}
 \nonumber   W=\frac{1}{32\pi^2} \int d^4 x\sqrt{g}\Bigg[\frac{1}{2}\left(\frac{1}{6}-\xi\right)^2 R(x) \Gamma(\Box)  R(x)+\\+ \frac{1}{180} R_{\mu\nu\alpha\beta}(x)\Gamma( \Box) R^{\mu\nu\alpha\beta}(x) -\frac{1}{180}  R_{\mu\nu}(x)\Gamma( \Box) R^{\mu\nu}(x)\Bigg]+O(R^3),
\end{gather}
where: 
\begin{align}
        \Gamma(\Box)=1-\log\left(\frac{\Box-i\epsilon}{\mu^2}\right).
    \end{align}
For grater details see \cite{Dobado:1998mr}.
Therefore the non-local term in the effective action produces an imaginary part of the form:

\begin{gather}
  Im \, W = \\=\nonumber \frac{1}{2}\frac{1}{1920\pi} \int \frac{d^4 p}{(2\pi)^4} \theta(-p^2) \left[C_{\mu\nu\alpha\beta}(p) C^{\mu\nu\alpha\beta}(-p) + 60 \left(\xi-\frac{1}{6}\right)^2 R(p)R(-p)\right]+O(R^3).
\end{gather}
For a static space-time the curvature tensor depends only on the spatial coordinates. Hence, the nonlocal term does not generate an imaginary part, since:
\begin{align}
Im\,  \log(\vec{p}^2-i\epsilon)=0.
\end{align}
(No branching of the logarithm on the real positive semiaxis.)
Thus according to these observations in static space-times there should not be any particle creation, at least not at the present level of approximation.  This point should be considered separately with greater care. 

For the metrics of the FLRW type the curvature tensor depends only on time. Therefore: 
\begin{gather}
   Im \, \int d^4 x \sqrt{g} R(x_0) \log(\Box-i\epsilon) R(x_0)
    \sim \\ \sim \nonumber 
   Im \, \int d^4 x \int d k_0 d p_0 e^{i (p_0+k_0)x_0} \sqrt{g} R(k_0) \log(-p^2_0-i\epsilon) R(p_0) 
    \sim \\ \nonumber \sim 
    V \, Im \, \int  d p_0  R(-p_0) \log(-p^2_0-i\epsilon) R(p_0)  \sim \\ \nonumber \sim  
    -\pi V \int  d p_0  R(-p_0) R(p_0)\sim \int d^4 x R(x_0)R(x_0).
\end{gather}
As a result for the FLRW metrics the imaginary part is given by:

\begin{gather}
  Im \, W = \frac{1}{2}\frac{1}{1920\pi} \int d^4 x \sqrt{g}    \left[C_{\mu\nu\alpha\beta}(x) C^{\mu\nu\alpha\beta}(x) + 60 \left(\xi-\frac{1}{6}\right)^2 R^2(x)\right]+O(R^3).
\end{gather}
But the FLRW metric is locally conformal to the Minkowskian metric and, hence, its Weyl tensor vanishes. Then, the imaginary part in such a case reduces to:
\begin{gather}
  Im \, W = \frac{1}{32\pi} \int d^4 x \sqrt{g}    \left[  \left(\xi-\frac{1}{6}\right)^2 R^2(x)\right]+O(R^3).
\end{gather}
For the case of conformal coupling it obviously vanishes.
    
\section{Conclusions (the relation to the method proposed by Wondrak et al.)}

In \cite{Wondrak:2023zdi} using the heat kernel expansion the effective action for the Euclidean metric is found to be: 
 \begin{align}
     W_E=-\frac{1}{2}\mu^{2 z} \int_0^\infty \frac{ds}{s^{1-z}}Tr\left(e^{-s(\triangle+\xi R)}\right) e^{-s(m^2-i\epsilon)}.
 \end{align}
Its imaginary part is defined as follows:
\begin{align}
     Im \, W_E = - Im\, \left( \frac{1}{32\pi^2}\mu^{2z} \int d^4 x\sqrt{g_E}\Gamma(z)(m^2-i\epsilon)^{-z}\frac{1}{180}\left[R_{\mu\nu\alpha\beta}(x) R^{\mu\nu\alpha\beta}(x) -   R_{\mu\nu}(x) R^{\mu\nu}(x) \right]\right).
\end{align}
In all, the authors of \cite{Wondrak:2023zdi} take only the imaginary part of the local terms and assume that curvature terms are real (using the Euclidean signature from the very beginning). Namely, using the following relation:

  \begin{align}
      \lim_{z\to 0,m^2\to 0}\, Im\, \left[\Gamma(z)(m^2-i\epsilon)^{-z}\right]=\lim_{m\to 0} \, Im\, \left[ \ln(m^2-i\epsilon)\right]=\pi \lim_{m\to 0}\theta(-m^2)=\frac{\pi}{2},
  \end{align}
they find the imaginary part of the effective action to be as follows:

\begin{align}
     Im \, W_E=- \frac{1}{64\pi} \int d^4 x\sqrt{g_E}\frac{1}{180}\left[R_{\mu\nu\alpha\beta}(x) R^{\mu\nu\alpha\beta}(x) -   R_{\mu\nu}(x) R^{\mu\nu}(x) \right].
 \end{align}
In terms of the Weyl tensor it has the form: 
 
\begin{align}
     Im \, W_E =- \frac{1}{4}\frac{1}{1920\pi} \int d^4 x\sqrt{g_E}C_{\mu\nu\alpha\beta}(x) C^{\mu\nu\alpha\beta}(x) .
 \end{align}
Which agrees with \eqref{WperC} up to a missing factor of two
\footnote{
This factor of two difference is still a question for us, but due to the equivalence of the methods we think that the answers should coincide and the discrepancy will be resolved in one way or another.}.
  
Thus, in \cite{Wondrak:2023zdi} it is claimed that the imaginary part can be obtained from the local terms of the heat kernel expansion. That is, they neglect non-local contributions from $R\log(\Box)R$ and etc.. Moreover, they consider the corresponding pair-creation rate as an analogue of Schwinger's expression \eqref{schwinger}.

We see two problems with this interpretation.

First, Schwinger's calculation involves an arbitrary number of interactions with the external field, and is even strongly non-perturbative in the sense that it does not yield an imaginary part if truncated to a finite number of interactions. The one by Wondrak et al. remains at the level of two interactions with the gravitational field, and therefore should be rather compared with the lowest-order perturbative pair-creation rate calculation given, for example, in the textbook \cite{Itzykson:1980rh}. However, as such it does not essentially go beyond the results of \cite{Frieman:1985fr}.

Second, by calculating directly in the Euclidean signature they are missing the Lorentzian constraint $\theta(-p^2)$. As was already pointed out by \cite{Ferreiro:2023jfs}, in the electromagnetic case this means loosing the information that a purely magnetic field is not able to pair-create (either perturbatively or non-perturbatively). Here the omission of the constraint leads to an analogous loss of information, which appears to be that gravitational pair creation is not possible in static spacetimes. 

However, in this respect the situation in the gravitational case is more serious since, if the metric depends on time, there is no straightforward way of the analytical continuation from the Euclidean signature to the Minkowskian one and back. In fact, in such a case the curvature tensor becomes complex-valued. (Instead, in symmetric backgrounds, such as de Sitter and anti de Sitter, one analytically continues in the complex plane of the geodesic or hyperbolic distance, rather than in the somplex time plane.) 

Let us close with two remarks on the massless case. First, in this limit besides the scalar pair creation by the gravitational field (that is by off-shell gravitons) also the inverse process of graviton pair creation by a scalar becomes possible in collinear kinematics, as well as the process of graviton to graviton and scalar \cite{adler:2016peh}. 
Second, in the electromagnetic case it is known (see, e.g., \cite{Dittrich:2000zu}) that Schwinger's formula \eqref{schwinger} cannot be applied in the massless case, since 
loop perturbation theory breaks down in that limit. Whether this caveat carries over to the gravitational case is presently unknown.

{\bf Acknowledgments:} C.S. thanks Anton Ilderton and Silvia Pla for correspondence. The work DVD was partially supported by the grant from the Foundation for the Advancement of Theoretical Physics and Mathematics ``BASIS''. The work of AET and DVD was partially supported by the Ministry of Science and Higher Education of the Russian Federation (agreement no. 075–15–2022–287).

 \bibliographystyle{unsrturl}
\bibliography{bibliography.bib}
 
 \end{document}